\newcounter{myctr}
\def\myitem{\refstepcounter{myctr}\bibfont\noindent\ifnum\themyctr>9\else\phantom{0}\fi\hangindent17pt\themyctr.\enskip}

\documentclass{ws-ijqi}
\usepackage{hyperref}
\usepackage[super,sort,compress]{cite}

\usepackage{dsfont}
\usepackage{braket}
\usepackage{color}
\usepackage{mathtools}
\usepackage{slashbox}
\usepackage{bbm}
\usepackage{subfigure} 
\usepackage{subcaption}
\allowdisplaybreaks

\newcommand{\be}{\begin{equation}}
\newcommand{\ee}{\end{equation}}
\newcommand{\ba}{\begin{eqnarray}}
\newcommand{\ea}{\end{eqnarray}}

\newtheorem{cor}{Corollary}

\newcommand{\one}{\mathds{1}}
\begin{document}

%%%%%%%%%%%%%%%%%%%%% Publisher's Area please ignore %%%%%%%%%%%%%%
%\catchline{}{}{}{}{}
%%%%%%%%%%%%%%%%%%%%%%%%%%%%%%%%%%%%%%%%%%%%%%%%%%%%%%%%%%%%%%%%%%%

\title{Secure key distribution based on Popescu-Rohrlich box fraction of dimensionally restricted nonlocality}

\author{Chellasamy Jebarathinam}

\address{Physics Division, National Center for Theoretical Sciences,\\ National Taiwan University,\\ Taipei 106319, Taiwan}

\maketitle

%\begin{history}
%\received{Day Month Year}
%\revised{Day Month Year}
%\accepted{Day Month Year}
%\comby{(xxxxxxxxxx)}
%\end{history}

\begin{abstract}
For the bipartite Bell scenario with two inputs and two outputs, a nonlinear witness of dimensionally restricted nonlocality is introduced.  Popescu-Rohrlich (PR) box fraction of dimensionally restricted nonlocality is then introduced and studied using the aforementioned witness and a nonlinear measure of correlations. This PR box fraction is also nonzero for certain Bell-local correlations. It is shown that any nonsignaling correlation shared by Alice and Bob that has dimensionally restricted nonlocality contains secrecy against any third party, Eve, who is also dimensionally restricted. In this context, for the specific Bell scenario considered,  it is demonstrated that the PR box fraction of dimensionally restricted nonlocality can be used as a resource for secure quantum key distribution, even if entanglement is not certified.    \end{abstract}


%\tableofcontents  % optional

\markboth{C. Jebarathinam}
{Secure key distribution based on dimensionally restricted nonlocality}

\section{Introduction} 
The nonlocality of quantum mechanics (QM), as demonstrated by the violation of a Bell inequality~\cite{Bel64}, can also be studied in the framework of generalized nonsignaling theories~\cite{BLM+05, MAG06}. This framework of quantum nonlocality underpins black-box information processing tasks~\cite{BCP+14}, such as device-independent quantum key distribution (QKD)~\cite{BHK05, AGM06}. 

In the framework of generalized nonsignaling theories, bipartite states are not given by vectors in a Hilbert space but by bipartite joint probability distributions; i.e. probabilities of a pair of results (outputs) given a pair of measurements (inputs). In other words, quantum correlations will be
replaced by more general “boxes” ( i.e. input-output devices).
Here, we shall focus on the simplest possible scenario,
namely, the case of two possible measurements for each party
(inputs $x, y \in \{0, 1\}$); each measurement providing a binary
result (outputs $a, b \in \{0, 1\}$). In this case, a bipartite state or correlation $P(ab|xy)$ is thus described by a set of $16$ joint probabilities.

The set of Bell-local distributions, which can be produced by using a classical state shared between the parties (a.k.a. local-hidden-variable (LHV) or shared randomness), forms a polytope, which has $16$ vertices (called deterministic distributions),
\begin{equation}
P^{\alpha\beta\gamma\epsilon}_D(ab|xy)=\left\{
\begin{array}{lr}
1, & a=\alpha x\oplus \beta\\
   & b=\gamma y\oplus \epsilon \\
0 , & \text{otherwise}.\\
\end{array}
\right.   
\label{eq:locdet}
\end{equation}   
Here, $\alpha,\beta,\gamma,\epsilon\in  \{0,1\}$ and  $\oplus$ denotes
addition modulo  $2$.   The local polytope is itself embedded in a
larger polytope, the nonsignaling polytope, which contains all states compatible with the nonsignaling principle. It
has $8$ nonlocal vertices (called Popescu-Rohrlich (PR)
boxes), 
\begin{align}
&P^{\alpha\beta\gamma}_{PR}(ab|xy)=\left\{
\begin{array}{lr}
\frac{1}{2}, & a\oplus b=x\cdot y \oplus \alpha x\oplus \beta y \oplus \gamma\\ 
0 , & \text{otherwise}\\
\end{array}
\right. \label{NLV}
\end{align}
which are all symmetries of the PR
box, $P_{PR}=P^{000}_{PR}$~\cite{PR94}. The set of states attainable by QM also forms a convex body, although not a polytope. The quantum set is strictly larger than the local polytope – quantum correlations can be Bell nonlocal – but strictly smaller than the nonsignalling polytope.

Any Bell-local distribution in the given scenario satisfies the complete set of
Bell-type inequalities~\cite{WW01a}, on the other hand, any given Bell nonlocal distribution violates one of these inequalities.  For the simplest scenario,  the  Clauser-Horne-Shimony-Holt (CHSH)   inequality~\cite{CHS+69} and its symmetries, which are given by
\begin{align}
&\mathcal{B}_{\alpha\beta\gamma} := (-1)^\gamma\braket{A_0B_0}+(-1)^{\beta \oplus \gamma}\braket{A_0B_1}\nonumber\\
&+(-1)^{\alpha \oplus \gamma}\braket{A_1B_0}+(-1)^{\alpha \oplus \beta \oplus \gamma \oplus 1} \braket{A_1B_1}\le2, 
\label{BCHSH}
\end{align}
form the  complete set,  where $\braket{A_xB_y}=\sum_{ab}(-1)^{a\oplus
  b}P(ab|xy)$.  
Quantum correlations violate the
CHSH inequality up to $2\sqrt{2}$~\cite{Cir80}.

In~\cite{Bie16}, it has been shown that any nonsignaling distribution
can be decomposed as a convex mixture of a single PR box and $16$ deterministic distributions. The PR box fraction of this decomposition, minimized overall possible decompositions, provides a measure of Bell nonlocality~\cite{EPR92, AGM06, SGB+06, Sca08, Vicente}, known as nonlocal cost $C(P)$ ~\cite{BCS+11}.

Bell nonlocality of the given any quantum correlations is shown against any LHV model of the correlation, i.e., $P(ab|xy)= \sum_\lambda p_\lambda P(a|x,\lambda) P(b|y,\lambda)$, where $\lambda$ denotes a classical state with probability distribution $\{p_\lambda\}_\lambda$, with $p_\lambda \ge 0$ and $\sum_\lambda p_\lambda =1$. Here, the LHV models are expressed using a discrete-variable classical state $\lambda$ (a.k.a finite shared randomness~\cite{BHQ+15}), which could also have been a continuous-variable state. In other words, Bell nonlocality is shown against any classical state $\lambda$ whose dimension $d_\lambda$, i.e., the number of values it takes in the LHV model, is unlimited.  Thus, quantum Bell nonlocality implies nonlocality independently of the dimension of the state used inside the box. This property of quantum Bell nonlocality was utilized to demonstrate secure QKD protocols in a device-independent way, i.e., independently of the dimension of the quantum state shared by the parties~\cite{AGM06,SGB+06,ABG+07}.

 \begin{figure}[t!]
\begin{center}
\includegraphics[width=7cm]{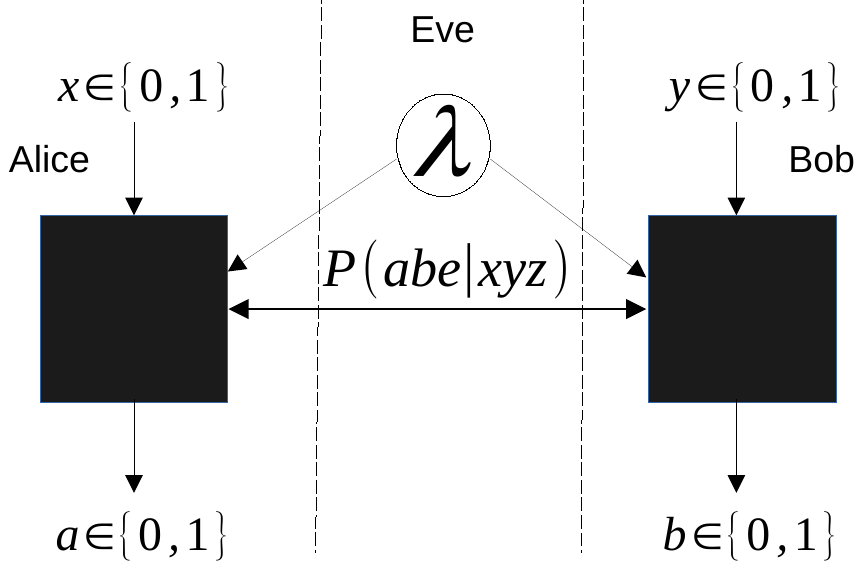}
\end{center}
\caption{The CHSH-protocol of secure key distribution in the device-independent scenario~\cite{AGM06}.  
\label{DIQKD}}
\end{figure}

In~\cite{AGM06}, the authors presented the CHSH-protocol of secure key distribution based on the PR box fraction of the nonlocal cost.  The CHSH-protocol consists of Eve, who has a tripartite probability distribution $P(abe|xyz)$, sending the marginal bipartite distribution $\sum_e P(abe|xyz)$ to Alice and Bob as in Fig.~\ref{DIQKD}. If the correlation shared by Alice and Bob has an LHV model, then the correlation can be shared with Eve by using a classical state $\lambda$. On the other hand, if the correlation has Bell nonlocality, then the correlation contains secrecy~\cite{MAG06}.   

 In this work, we consider dimensionally restricted nonlocality (DRNL), i.e., nonlocality shown against any classical state $\lambda$ whose dimension is limited to the number of measurement results. 
 Suppose a correlation $P(ab|xy)$ is produced using a bipartite quantum state, with local dimensions $d_A$ and $d_B$ on Alice's and Bob's side, respectively,  with the number of results $n=\min\{d_A,d_B\}$ on each side. Then the correlation  has DRNL if it does not have a dimensionally restricted LHV model, i.e.,
\begin{align} \label{BLd<=d_o}
P(ab|xy)= \sum^{d_{\lambda^*}-1}_{{\lambda^*}=0} p_{\lambda^*} P(a|x,{\lambda^*})P(b|y,{\lambda^*}). 
\end{align}
Here, $\lambda^*$ denotes a classical state that minimizes $d_\lambda$ required to simulate the correlation and satisfies $d_{\lambda^*} \le n$.
The LHV model of the given correlation as a convex mixture of local deterministic states does not necessarily imply that the finite shared randomness in this context minimizes $d_\lambda$. For instance,  in the case of the CHSH scenario, $d_{\lambda^*} \le 4$~\cite{DW05}, at the same time, there exist Bell-local distributions which require $d_\lambda >4$ if a deterministic distribution is used for each $\lambda$~\cite{JAS17}.  

DRNL can also be observed by certain Bell-local quantum correlations ~\cite{Jeb14,JAS17,JD23}. Any such Bell-local correlation has superlocality~\cite{DW05}, i.e.,  $d_\lambda{^*}>\min\{d_A,d_B\}$. Superlocality can be interpreted as a means of certifying quantum discord~\cite{OZ01,HV01} by assuming only the Hilbert space dimension of Alice and Bob ~\cite{JAS17,JD23,JKC+24}.

In the following, we demonstrate a way to detect and quantify DRNL.
Based on this characterization of DRNL, we then demonstrate how a secure QKD can be achieved.

\section{Characterizing DRNL}
To detect DRNL, we consider a nonlinear determinant witness in terms of  
the covariance of  $A_x$ and $B_y$ given by  $\text{cov}(A_x,B_y)=\braket{A_xB_y} -\braket{A_x}\braket{B_y}$. $\braket{A_x}$ and $\braket{B_y}$ are marginal expectation values. The nonlinear witness denoted by ${NL}$ is given by
\begin{align}
NL=\left|\begin{array}{cc}\text{cov}(A_0,B_0) & \text{cov}(A_0,B_1)\\ 
\text{cov}(A_1,B_0) & \text{cov}(A_1,B_1) \end{array}\right|. \label{Wit}
\end{align}
A nonzero value of $NL$ witnesses DRNL. To show this, we obtain the following proposition. 
\begin{proposition}
For any Bell-local distribution that can be simulated by any classical state $\lambda$ having dimension $d_\lambda \le 2$, $NL=0$. 
\label{propo01}
\end{proposition} 
The proof of this statement is given in~\ref{GWDRNL}. 
From Proposition~\ref{propo01}, it follows that $NL$ can be used to witness DRNL since $NL>0$ implies $d_\lambda* \nleq 2$.  The nonlinear  witness $NL$ is inspired by the nonlinear determinant witness introduced in~\cite{BQB14} as a dimension witness in a prepare-and-measure scenario, and, subsequently, shown in~\cite{JD23}, as a witness of superlocality of two-qubit states.  In~\cite{QYY25}, the nonlinear witness (\ref{Wit}) was studied as a witness of quantum discord with uncharacterized devices.  

Next, to introduce a measure of DRNL analogous to the nonlocal cost of Bell nonlocality, consider a specific family of nonsignaling correlations by the noisy PR box, $P^{\alpha\beta\gamma}_{nPR}$, which is a mixture of a PR box and white noise,
\begin{equation} 
P^{\alpha\beta\gamma}_{nPR}=p_{PR}P^{\alpha\beta\gamma}_{PR}+(1-p_{PR})P_N,
\label{PRiso} 
\end{equation} 
where $p_{PR}$ satisfies $0\le p_{PR}\le 1$ and  $P_N$ is the  maximally mixed state,  i.e., $P_N(ab|xy)=1/4$  for  all  $x,y,a,b$.
For $p_{PR} \le 1/\sqrt{2}$, the noisy PR box can
be distributed by QM (e.g., by a Werner state,
$\rho_W=WP_{\phi^+}+(1-W)\one/4$, with visibility $W=\sqrt{2}p_{PR}$, where
$P_{\phi^+}$ denotes the projector onto
$\ket{\phi^+}=(\ket{00}+\ket{11})/\sqrt 2$).
The noisy PR box violates a CHSH inequality, that is, $\mathcal{B}_{\alpha\beta\gamma}=4p_{PR}>2$ if and only if  $p_{PR}>\frac{1}{2}$ and has the nonlocal cost $C(P^{\alpha\beta\gamma}_{nPR})=\min\{0,2p_{PR}-1\}$.
On the other hand, the witness of DRNL (\ref{Wit}) takes the value given by $NL(P^{\alpha\beta\gamma}_{nPR})=2p^2_{PR}$. This implies that the noisy PR box within QM has DRNL for any $p_{PR}>0$.
%Note that when the noisy PR box is Bell-local, DR nonlocality certifies entanglement if $p_{PR} > \frac{1}{2\sqrt{2}}$~\cite{GBS16}, on the other hand, quantum discord is certified for any $p_{PR}>0$~\cite{JAS17,JD23}.    

Note that for the noisy PR box (\ref{PRiso}), $NL(P^{\alpha\beta\gamma}_{nPR})>0$ if and only if the PR box fraction $p_{PR}>0$. Given any nonsignaling distribution, the single PR box fraction   that indicates DRNL is called the PR box fraction of DRNL.
To define the PR box fraction of DRNL beyond the noisy PR box and capture it using a measure of correlations, we consider a quantity, denoted by $\Gamma$, constructed in terms of the covariance CHSH inequalities~\cite{PHC+17}. 
Define  the  absolute covariance CHSH functions  $\text{cov}\mathcal{B}_{2\alpha+\beta}  :=
| \text{cov}(A_0B_0)        +         (-1)^{\beta} \text{cov}(A_0B_1)        +
(-1)^{\alpha} \text{cov}(A_1B_0)  + (-1)^{\alpha  \oplus  \beta \oplus  1}
 \text{cov}(A_1B_1)|$. Consider the following triad of quantities constructed from these four covariance CHSH functions:
\begin{align}  
\begin{split}
\Gamma_1&:= \Big||\text{cov}\mathcal{B}_0  - \text{cov}\mathcal{B}_1  | -
|\text{cov}\mathcal{B}_2  - \text{cov}\mathcal{B}_3|\Big|\\
\Gamma_2&:= \Big||\text{cov}\mathcal{B}_0  - \text{cov}\mathcal{B}_2  | -
|\text{cov}\mathcal{B}_1  -\text{cov}\mathcal{B}_3|\Big| \\
\Gamma_3&:= \Big||\text{cov}\mathcal{B}_0  - \text{cov}\mathcal{B}_3  | -
|\text{cov}\mathcal{B}_1  - \text{cov}\mathcal{B}_2|\Big|.\label{gi}
\end{split}
\end{align}
Here, each $\Gamma_i$ is constructed such that $\Gamma_i \ge 0$ due to the outer $|\cdot|$ in Eq. (\ref{gi}), which takes the absolute value of the quantity within it; $\Gamma_i=0$ for any product distribution of the form $P(ab|xy)=P(a|x)P(b|y)$ due to $\text{cov}\mathcal{B}_{2\alpha+\beta}=0$ for any product distribution;  $\Gamma_i=4$, which is the algebraic maximum of $\Gamma_i$ within all nonsignaling correlations, is achieved by any PR box $P^{\alpha\beta\gamma}_{PR}$ (since for any PR box only one of the covariance CHSH $\text{cov}\mathcal{B}_{2\alpha+\beta}$ takes its algebraic maximum of $4$ and others are zero).  $\Gamma$ is then defined:
\begin{equation}
\Gamma:= \min_i \Gamma_i. 
\label{eq:G}
\end{equation}
Here, $\Gamma$ is constructed using the three $\Gamma_i$ in Eq. (\ref{gi}) so that it is invariant under relabeling of inputs and/or outputs \footnote{In~ \cite{Jeb14}, a measure called Bell discord (also appeared as Bell strength in~\cite{JAS17}), which is the same as the quantity $\Gamma$ defined in Eq. (\ref{eq:G}), but without invoking covariance on the CHSH functions, was studied. This measure was considered in~\cite{Jeb14} to capture the maximal irreducible PR box fraction in any convex mixture of the PR boxes, which can be Bell nonlocal or not.}. This follows because it can be checked that if Alice performs the operations $x \rightarrow x \oplus 1$ and/or $a \rightarrow a \oplus \alpha x \oplus \beta$,  $\Gamma$ is invariant due to each $\Gamma_i$ is invariant or one of $\Gamma_i$ is transformed into another $\Gamma_{i'}$ under such operations. Similarly, it can be checked that if Bob performs the relabeling of inputs and/or outputs, 
$\Gamma$ is invariant.
Thus, $\Gamma$ satisfies the following axiomatic properties of being a measure of correlations: (i) $ 0 \le \Gamma \le 4$; (ii) $\Gamma = 0$ for any product distribution of the form, $P(ab|xy)=P(a|x)P(b|y)$; (iii)  $\Gamma = 4$ for any PR box $P^{\alpha\beta\gamma}_{PR}$; and (iv) $\Gamma$ is invariant under relabeling of inputs and/or outputs. 

Note that for the noisy PR box (\ref{PRiso}), $\Gamma(P^{\alpha\beta\gamma}_{nPR})=4p_{PR}>0$ if and only if the PR box fraction of DRNL is nonzero. Thus, for the specific family of
nonsignaling distributions,  a nonzero $\Gamma$ indicates the PR box fraction of DRNL as a measure of DRNL.  

Moving beyond the noisy PR box, a nonzero $\Gamma$ does not, in general, imply DRNL. This follows because certain Bell-local correlations with $d_{\lambda}\le 2$ also have $\Gamma>0$ (see \ref{App:G>0} for the illustration). However, in the following, we demonstrate that  $\Gamma$ captures the PR box fraction of DRNL for a subset of all nonsignaling distributions going beyond the noisy PR box. To this end, we first obtain the following lemma. 
\begin{lemma}\label{lem2}
Any convex combination of eight PR boxes $\sum_{\alpha\beta\gamma} p_{\alpha\beta\gamma} P^{\alpha\beta\gamma}_{PR}$ can be reexpressed as  
\begin{equation} \label{cm8PR}
\sum_{\alpha,\beta,\gamma} p_{\alpha\beta\gamma} P^{\alpha\beta\gamma}_{PR}=p_{NL} P_{NL} + (1-p_{NL}) P_{\Gamma=0}, 
\end{equation}
where  $4p_{NL}=\Gamma(\sum_{\alpha,\beta,\gamma} p_{\alpha\beta\gamma} P^{\alpha\beta\gamma}_{PR})$, $P_{NL}$ is one of the PR boxes and  $P_{\Gamma=0}$ is a Bell-local distribution with $\Gamma=0$.  $P_{\Gamma=0}$ is a convex mixture of seven Bell-local distributions, which are the uniform mixture of two PR boxes. 
\end{lemma} 
The proof of this lemma is given in~\ref{dconPR}. 
Next, using Lemma~\ref{lem2}, we obtain the following theorem.
\begin{theorem}\label{thm}
Consider any nonsignaling distribution $P$ that can be decomposed as a convex mixture of a single PR-box and a Bell-local distribution as follows:
\begin{equation} \label{PRdecom} 
P=p_{PR} P^{\alpha\beta\gamma}_{PR} +  (1-p_{PR}) P_{\Gamma=0}
\end{equation}
  where $P_{\Gamma=0}$ is a Bell-local distribution with $\Gamma=0$, which need not be as in Eq.~(\ref{cm8PR}). For any such $P$ given above, $\Gamma(P)=4p_{PR}$. 
\end{theorem}
\begin{proof}
We express any nonsignaling distribution $P$ as a convex mixture of eight PR boxes and a Bell-local distribution as follows:
\begin{align}\label{step1} 
P&=\sum_{\alpha,\beta,\gamma} p_{\alpha\beta\gamma} P^{\alpha\beta\gamma}_{PR} + \left(1-\sum_{\alpha\beta\gamma} p_{\alpha\beta\gamma}\right) P_L. 
\end{align}  
For over all decompositions of this form, we choose a particular decomposition for which $P_{L}$ cannot be expressed in terms of the PR boxes with a nonzero weight for at least one of the PR boxes, i.e., 
\begin{align}\label{step2} 
P&=\sum_{\alpha,\beta,\gamma} p'_{\alpha\beta\gamma} P^{\alpha\beta\gamma}_{PR} + \left(1-\sum_{\alpha\beta\gamma} p'_{\alpha\beta\gamma}\right)P_{PR_\downarrow}. 
\end{align} 
Here, $P_{PR_\downarrow}$ can only be expressed as a convex mixture of deterministic states and it has $\Gamma \ge 0$. 

Let us now restrict ourselves to any $P$ that can be expressed as 
\begin{align}\label{step3} 
P&=\sum_{\alpha,\beta,\gamma} p'_{\alpha\beta\gamma} P^{\alpha\beta\gamma}_{PR} + \left(1-\sum_{\alpha\beta\gamma} p'_{\alpha\beta\gamma}\right)P^{\Gamma=0}_{PR_\downarrow}, 
\end{align} 
where $P^{\Gamma=0}_{PR_\downarrow}$ cannot be expressed in terms of the PR boxes with a nonzero weight for at least one of the PR boxes and has $\Gamma=0$.
For any $P$ as given by Eq.~(\ref{step3}),
\be \label{step4}
\Gamma(P)=\Gamma(\sum_{\alpha,\beta,\gamma} p'_{\alpha\beta\gamma} P^{\alpha\beta\gamma}_{PR}).
\ee
This follows because if $P$ in Eq.~(\ref{step3}) has a nonzero $\Gamma$, then the nonzero $\Gamma$ is isolated in the first part of the decomposition in Eq.~(\ref{step3}).

From Lemma~\ref{lem2}, it follows that any convex mixture of the eight PR boxes in the decomposition (\ref{step3}) can be reexpressed as a convex mixture of a single PR box and a Bell-local distribution with $\Gamma=0$ as in Eq.~(\ref{cm8PR}).
Using this decomposition for the mixture of the eight PR boxes in Eq. (\ref{step3}), we obtain 
\begin{align} \label{step5}
P&= p_{NL} P_{NL} + p_{\Gamma=0}P_{\Gamma=0}+(1-p_{NL}-p_{\Gamma=0})P^{\Gamma=0}_{PR_\downarrow}, 
\end{align}
where $4p_{NL}=\Gamma(P)$ using Eq.~(\ref{step4}) and $p_{NL}+p_{\Gamma=0}=\sum_{\alpha,\beta,\gamma} p'_{\alpha\beta\gamma}$.
We reexpress $P$ as given above as follows:
\begin{align} \label{step6}
P&= p_{NL} P_{NL} + (1-p_{NL})P_L, 
\end{align}
where 
\be \label{step7}
P_L=\frac{1}{1-p_{NL}}(p_{\Gamma=0}P_{\Gamma=0}+(1-p_{NL}-p_{\Gamma=0})P^{\Gamma=0}_{PR_\downarrow}).
\ee
Now, note that $P_L$ in Eq.~(\ref{step6}) must have $\Gamma=0$. This follows because if $p_{NL}=0$ in Eq.~(\ref{step6}), then $\Gamma(P)=0$ since $\Gamma(P)=4p_{NL}$. This holds if $P_L$ has $\Gamma=0$.
\end{proof}
Let us now illustrate an usefulness of the above theorem. To this end, note that moving beyond the noisy PR box, a nonzero $NL$ does not, in general, imply a nonzero value for the PR box fraction of DRNL. This follows because certain Bell-local correlations with $d_{\lambda}> 2$ also have $NL>0$, without having a fraction of PR box as in the case of the noisy PR box.  For example, 
the Bell-local correlation,
\be
P=\frac{1}{2}(P^{000}_{PR}+P^{111}_{PR}),
\ee
has $NL=1$, at the same time, it does not have the PR box fraction of DRNL, which can be seen by noting that it has $\Gamma=0$.
Thus, for any given $P$ which has $NL>0$ and admits a decomposition as in Eq.~(\ref{PRdecom}), $\Gamma$ serves as  a quantifier of DRNL by the PR box fraction. While we have been able to characterize the PR box fraction of DRNL for a broader class of correlations using Theorem~\ref{thm},  it remains a challenge to provide a complete characterization of the PR box fraction of DRNL.

\section{Secure key distribution using DRNL as a resource}
\begin{figure}
	\begin{subfigure}
		\centering\includegraphics[width=7cm]{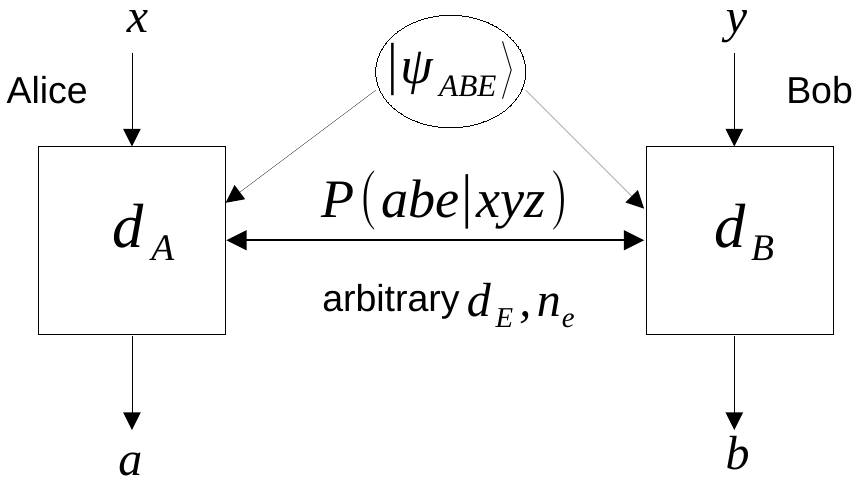}
		\caption{Entanglement-based QKD scenario, where Eve holds a purification $\ket{\psi_{ABE}}$ of the state shared by Alice and Bob $\rho_{AB}$. Here, local dimensions $d_A$ and $d_B$ of Alice and Bob are trusted, while Eve's dimension $d_E$ is arbitrary, which implies that the number of measurement results of Eve could be $n_e \le d_E$. %In this scenario, the secret correlations between Alice and Bob are certified by verifying entanglement using standard entanglement witnesses \cite{AG05} or measurement-device-independent entanglement witnesses \cite{GBS16}.    
        \label{Fig:SQKD}}
	\end{subfigure}
	\begin{subfigure}
		\centering\includegraphics[width=7cm]{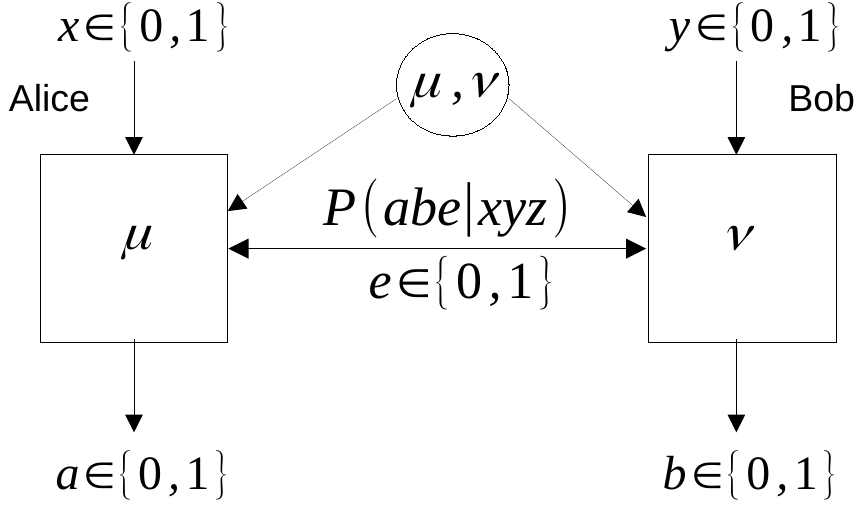}
		\caption{The CHSH-protocol of secure key distribution based on DRNL. Here, Eve is also dimensionally restricted, and the randomness of Eve is described by local randomness $\mu$ and $\nu$, which are uncorrelated.\label{Fig:DRQKD}}
	\end{subfigure}
	\end{figure}

We shall study the secrecy of DRNL to achieve a secure quantum key distribution in the context of the CHSH scenario.
Before that, consider observing secret correlations for QKD scenarios as in Fig.~\ref{Fig:SQKD} based on entanglement witnesses ~\cite{AG05}, which require trust on the quantum measurements, or certifying entanglement in a measurement-device-independent way in which measurements are uncharacterized, but local dimensions are characterized ~\cite{GBS16}. In this context, Eve has a purification $\ket{\psi_{ABE}}$ in $\mathcal{L}(\mathcal{H}_{d_A} \otimes \mathcal{H}_{d_B} \otimes \mathcal{H}_{d_E})$ of the state $\rho_{AB}$ in $\mathcal{L}(\mathcal{H}_{d_A} \otimes \mathcal{H}_{d_B})$. Here, $d_A$ and $d_B$ are characterized, while $d_E$ is arbitrary, which implies that the number of results of Eve's measurements $n_e$ in $P(abe|xyz)$   could be as large as $d_E$, i.e., $n_e \le d_E$. Thus, given that Eve's state is not dimensionally restricted, entanglement is required to produce secret correlations~\cite{AG05} in scenarios as in Fig.~\ref{Fig:SQKD}.
This motivates us to ask whether any correlation that has DRNL has secrecy if Eve is also dimensionally restricted. To answer this question, we show the following result.
\begin{theorem} \label{secretdrnl}
    Any correlation $P(ab|xy)$ that has DRNL contains secrecy against any Eve's state that is also dimensionally restricted.
\end{theorem}
\begin{proof}
If any given correlation $P(ab|xy)$ does not have DRNL, then the correlation can be shared with Eve's state, which has dimension as that of Alice or Bob, i.e., 
\begin{equation} \label{nsc}
P(abe|xyz)=\sum^{d_e-1}_{e=0} p_e P(a|x,e) P(b|y,e) P(e|z), 
\end{equation}
where $d_e \le \min\{d_A,d_B\}$. On the other hand, if Alice and Bob do not share secret correlation against any dimensionally restricted Eve, then the extension of the correlation to the tripartite probability distribution satisfies (\ref{nsc}). Hence, the correlation shared by Alice and Bob does not have DRNL.
 It then follows that any correlation that has DRNL has secrecy against Eve, who is also dimensionally restricted, since  $P(abe|xyz)\ne \sum_{e} p_e P(a|x,e) P(b|y,e) P(e|z)$.
\end{proof}

Next, we identify the relevant scenarios where secrecy of DRNL can be used for real-world applications. Note that if Eve distributes the state $\rho_{AB}$ shared by Alice and Bob, then DRNL is useful to establish secret correlation if entanglement is certified ~\cite{AG05,GBS16}. Our goal is to identify scenarios where DRNL is useful, even if entanglement is not certified. To this end, we consider a dimensionally restricted scenario where the state shared by Alice and Bob is not provided by Eve, but rather by a trusted source, for instance, Alice, who sends half of the state to Bob while keeping the other half with her. In this context, Eve's role is characterized by providing the measurement devices to Alice and Bob and having the extension $P(abe|xyz)$ of $P(ab|xy)$ observed by Alice and Bob.

Now, note that DRNL can be created from a state given by Eq.~(\ref{BLd<=d_o}) if Eve uses shared randomness $\lambda$ between Alice and Bob. This follows because any Bell-local correlation that has DRNL can be prepared by a convex mixture of distributions that do not have DRNL.  Thus, if DRNL is used as a resource, shared randomness $\lambda$ cannot be freely used between Alice and Bob. Therefore, we consider a scenario as in Fig.~\ref{Fig:DRQKD}, where the source of randomness by Eve is given by two uncorrelated internal randomness $\mu$ and $\nu$ acting on Alice and Bob, respectively, i.e., the joint probability distribution $p(\mu,\nu)$  satisfies $p(\mu,\nu)=q(\mu)r(\nu)$. %That is, $P(abe|xyz)$ is given by
 %$P(abe|xyz)=\sum_{\mu,\nu} q_\mu r_\nu P(abe|xyz,\mu,\nu)$, where $P(ab|xy)=\sum_{\mu,\nu} q_\mu r_\nu P(abe|xy,\mu,\nu)$. 
 Within QM, the scenario as in Fig.~\ref{Fig:DRQKD} provides a measurement-device-independent scenario to obtain secret correlation using any distribution that has DRNL, provided that Eve's state is also dimensionally restricted.

Before we proceed to study the key rate of the secret correlations obtained from the CHSH-protocol~\cite{AGM06} in the dimensionally restricted scenario as in Fig.~\ref{Fig:DRQKD}, we obtain the following corollary of  Theorem~\ref{secretdrnl}.
\begin{cor} \label{secdrnl}
    Any correlation $P(ab|xy)$ that has DRNL is completely uncorrelated with dimensionally restricted Eve, i.e., $P(abe|xyz)=P(ab|xy)P(e|z)$, where $P(ab|xy)$ has DRNL and $P(e|z)$  arises from a state which has dimensional restriction as that of Alice or Bob.
\end{cor}
\begin{proof}
From Bayes' theorem, $P(abe|xyz)=P(ab|xyze)P(e|xyz)$. Using nonsignaling from Alice and Bob to Eve, it then follows that 
\begin{equation} \label{bayt}
P(abe|xyz)=P(ab|xyze)P(e|z). 
\end{equation}
From nonsignaling from Eve to Alice and Bob, 
$P(ab|xy)=\sum_e P(abe|xyz)$ for all Eve's measurements $z$. Now, using Eq. (\ref{bayt}),
\begin{align}
P(ab|xy)=\sum_e P(ab|xye)P(e|z) \quad \forall \quad z.
\end{align}
Now, for any given $z$,
\begin{align}
P(ab|xy)=\sum_e P(e) P(ab|xye), 
\end{align}
where $e$ plays the role of shared randomness $\lambda$. 
Since shared randomness $\lambda$ cannot be freely used between Alice and Bob if DRNL is used as a resource,  we have $P(ab|xye)=P(ab|xy)$. Using this constraint in Eq.~(\ref{bayt}),  we have
\begin{equation} 
P(abe|xyz)=P(ab|xy)P(e|z). 
\end{equation}
\end{proof}

In the CHSH-protocol to obtain a shared secure key~\cite{AGM06}, Alice and Bob randomly input their bits into their respective devices and collect their output bits. Alice reveals her inputs, and Bob flips his output bit when both of their inputs are $1$. 
For one-way key distribution protocols in which communication goes from Alice to Bob, the secret bit shared by Alice and Bob is quantified by the one-way key rate $K^\rightarrow \ge I(A:B) - I(A:E)$~\cite{DW05}.      
If Alice and Bob share secret correlation in the PR box, $P_{PR}$, then the maximal secure key rate of $1$ bit is achieved against any Eve because $I(A:B)=1$ and $I(A:E)=0$ for the PR box, which is an extremal nonsignaling box. 

In QM, the PR box cannot be realized perfectly~\cite{RTH+16}. 
In~\cite{AGM06}, the secret key rate of the noisy PR box, $P_{nPR}:=p_{PR} P_{PR}+(1-p_{PR}) P_N$, achievable by the PR box fraction of the nonlocal cost, was studied. In this context, a positive value of the key rate against the individual attack of nonsignaling (supraquantum) Eve was shown if $C(P_{nPR}) > 0.414$

Since we consider the CHSH-protocol to establish a secure key distribution in a scenario as in Fig.~\ref{Fig:DRQKD}, the key rate is given by $K^\rightarrow \ge I(A:B)$, provided that the PR box fraction of DRNL is nonzero. This follows because $I(A:E)=0$ for any correlation with DRNL, which follows from Corollary~\ref{secdrnl} and the nonzero PR box fraction provides the key generation.  

We shall study the secure key rate of the noisy PR box $P_{nPR}$ with DRNL.  The secure key rate of the noisy PR box within QM, realized by the Werner state $\rho_W$, has been checked to be $K^\rightarrow \ge I(A:B)=1-h\left(\frac{1}{2}\left(1-\frac{W}{\sqrt{2}}\right)\right)$, where $h$ is the binary entropy. This key rate is nonzero for any Werner state, i.e., $W>0$, and its secrecy is implied by $\Gamma(P_{nPR})=4p_{PR}$, with $p_{PR}=W/\sqrt{2}$, being nonzero for any $W>0$.  Thus, using DRNL as a resource in the scenario as in Fig.~\ref{Fig:DRQKD}, the noisy PR box within QM is useful for secure key distribution for any $p_{PR} >0$. Note that when the noisy PR box is Bell-local, DRNL certifies entanglement if $p_{PR} > \frac{1}{2\sqrt{2}}$~\cite{GBS16}, on the other hand, quantum discord is certified for any $p_{PR}>0$~\cite{JAS17,JD23}. Thus, the noisy PR box is useful for secure key distribution even when entanglement is not certifiable for $0 < p_{PR} \le  \frac{1}{2\sqrt{2}}$.  

\section{Conclusions and Discussions}
For the bipartite Bell scenario with two inputs and two outputs, a nonlinear witness of DRNL has been shown. 
The PR box fraction of DRNL has been studied using a nonlinear measure of correlations constructed using covariance CHSH functions. It has been demonstrated that the PR box fraction of DRNL can be used to provide a secure QKD if Eve is also dimensionally restricted. On the other hand, in the presence of the most powerful Eve, who is not dimensionally restricted, secret correlations are present in DRNL if entanglement is certified~\cite{AG05,GBS16}.     

With quantum Bell nonlocality as a resource, the noises acting on the devices held by Alice and Bob  are modeled by shared randomness~\cite{LBF+23}. Whereas, with the PR box fraction of DRNL as a resource, the realization of secure QKD  with a noisy PR box models the noises as internal sources of randomness in Alice's and Bob's devices, which are uncorrelated, as modeled for selftesting quantum random number generation in~\cite{LBL+15}.   

QKD based on the PR box fraction of DRNL has practical advantages in the following real-world scenarios.  In the presence of inefficient detectors, Alice and Bob may not be able to observe Bell nonlocality genuinely~\cite{Bra11}, however, the LHV models produced in the experiment in observing Bell nonlocal correlations, can still be used to achieve secure QKD in the context of dimensionally restricted Bell scenario as in Fig.~\ref{Fig:DRQKD}~\cite{GBS16}. Given that the volume of the set of separable states is nonzero~\cite{KHS+98}, preserving entanglement of quantum systems becomes difficult in the presence of noise such as dissipation or decoherence~\cite{Zur03}. In this context, given that quantum discord can be preserved in the presence of noise~\cite{FAC+10,GLH+12}, the observation of the PR box fraction of DRNL is therefore relevant to achieve secure QKD.

\section*{Acknowledgement}
 This work was supported by the National Science and Technology Council, the Ministry of Education (Higher Education Sprout Project NTU-113L104022-1), and the National Center for Theoretical Sciences of Taiwan.

\appendix

\section{Proof of Proposition~\ref{propo01}}\label{GWDRNL}
Any Bell-local distribution, which can be simulated by any classical state $\lambda$ having dimension $d_\lambda \le 2$, can be expressed in terms of the deterministic distributions as follows: 
\begin{equation} \label{BLd<=2}
P_L(ab|xy)= \sum^1_{\lambda=0} p_\lambda P_\lambda(a|x)P_\lambda(b|y), 
\end{equation}
where $P_\lambda(a|x)$ and $P_\lambda(b|y)$ are a convex combination of deterministic local states given by   
\begin{align}
P_\lambda(a|x)&=\sum^3_{i=0} k_{i}^{(\lambda)} P^{i}_D(a|x),\\
P_\lambda(b|y)&=\sum^3_{j=0} l_{j}^{(\lambda)} P^{j}_D(b|y),
\end{align}
where  $P^{i}_D(a|x)$  and $P^{j}_D(b|y)$ are given by
\begin{equation}
P^{\alpha\beta}_D(a|x)=\left\{
\begin{array}{lr}
1, & a=\alpha x\oplus \beta\\
0 , & \text{otherwise},\\
\end{array}
\right.   
\label{eq:locdetA}
\end{equation} 
with $i=\alpha\oplus \beta$, and
\begin{equation}
P^{\gamma\epsilon}_D(b|y)=\left\{
\begin{array}{lr}
1, & b=\gamma x\oplus \epsilon\\
0 , & \text{otherwise},\\
\end{array}
\right.   
\label{eq:locdetB}
\end{equation}
with $j=\gamma \oplus \epsilon$, respectively.  For any Bell-local distribution as expressed in Eq.~(\ref{BLd<=2}), $\text{cov}(A_x,B_y)$ are given by
\begin{align}
\text{cov}(A_0,B_0)&=p_0 (k^{(0)}_0+k^{(0)}_2-k^{(0)}_1-k^{(0)}_3)(l^{(0)}_0+l^{(0)}_2-l^{(0)}_1-l^{(0)}_3) \nonumber \\
&+p_1 (k^{(1)}_0+k^{(1)}_2-k^{(1)}_1-k^{(1)}_3)(l^{(1)}_0+l^{(1)}_2-l^{(1)}_1-l^{(1)}_3) \nonumber \\
&-\{p_0 (k^{(0)}_0+k^{(0)}_2-k^{(0)}_1-k^{(0)}_3)+p_1 (k^{(1)}_0+k^{(1)}_2-k^{(1)}_1-k^{(1)}_3)\} \nonumber \\
&\times \{p_0 (l^{(0)}_0+l^{(0)}_2-l^{(0)}_1-l^{(0)}_3)+p_1 (l^{(1)}_0+l^{(1)}_2-l^{(1)}_1-l^{(1)}_3)\} \\
\text{cov}(A_0,B_1)&=p_0 (k^{(0)}_0+k^{(0)}_2-k^{(0)}_1-k^{(0)}_3)(l^{(0)}_0+l^{(0)}_3-l^{(0)}_1-l^{(0)}_2) \nonumber \\
&+p_1 (k^{(1)}_0+k^{(1)}_2-k^{(1)}_1-k^{(1)}_3)(l^{(1)}_0+l^{(1)}_3-l^{(1)}_1-l^{(1)}_2) \nonumber \\
&-\{p_0 (k^{(0)}_0+k^{(0)}_2-k^{(0)}_1-k^{(0)}_3)+p_1 (k^{(1)}_0+k^{(1)}_2-k^{(1)}_1-k^{(1)}_3)\} \nonumber \\
&\times \{p_0 (l^{(0)}_0+l^{(0)}_3-l^{(0)}_1-l^{(0)}_2)+p_1 (l^{(1)}_0+l^{(1)}_3-l^{(1)}_1-l^{(1)}_2)\} \\
\text{cov}(A_1,B_0)&=p_0 (k^{(0)}_0+k^{(0)}_3-k^{(0)}_1-k^{(0)}_2)(l^{(0)}_0+l^{(0)}_2-l^{(0)}_1-l^{(0)}_3) \nonumber \\
&+p_1 (k^{(1)}_0+k^{(1)}_3-k^{(1)}_1-k^{(1)}_2)(l^{(1)}_0+l^{(1)}_2-l^{(1)}_1-l^{(1)}_3) \nonumber \\
&-\{p_0 (k^{(0)}_0+k^{(0)}_3-k^{(0)}_1-k^{(0)}_2)+p_1 (k^{(1)}_0+k^{(1)}_3-k^{(1)}_1-k^{(1)}_2)\} \nonumber \\
&\times \{p_0 (l^{(0)}_0+l^{(0)}_2-l^{(0)}_1-l^{(0)}_3)+p_1 (l^{(1)}_0+l^{(1)}_2-l^{(1)}_1-l^{(1)}_3)\}  \\
\text{cov}(A_1,B_1)&=p_0 (k^{(0)}_0+k^{(0)}_3-k^{(0)}_1-k^{(0)}_2)(l^{(0)}_0+l^{(0)}_3-l^{(0)}_1-l^{(0)}_2) \nonumber \\
&+p_1 (k^{(1)}_0+k^{(1)}_3-k^{(1)}_1-k^{(1)}_2)(l^{(1)}_0+l^{(1)}_3-l^{(1)}_1-l^{(1)}_2) \nonumber \\
&-\{p_0 (k^{(0)}_0+k^{(0)}_3-k^{(0)}_1-k^{(0)}_2)+p_1 (k^{(1)}_0+k^{(1)}_3-k^{(1)}_1-k^{(1)}_2)\} \nonumber \\
&\times \{p_0 (l^{(0)}_0+l^{(0)}_3-l^{(0)}_1-l^{(0)}_2)+p_1 (l^{(1)}_0+l^{(1)}_3-l^{(1)}_1-l^{(1)}_2)\}.
\end{align}
Using these expressions of $\text{cov}(A_x,B_y)$, computing $NL$ gives $NL=0$. Therefore, $NL=0$ for any Bell-local distribution as given by Eq.~(\ref{BLd<=2}).

\section{\texorpdfstring{$\Gamma$}{Gamma} of classical states with \texorpdfstring{$d_{\lambda^*}\le 2$}{d-lambda* <= 2}}\label{App:G>0}
With the expressions of $\text{cov}(A_x,B_y)$ for any Bell-local distribution (\ref{BLd<=2}), calculating $\text{cov}\mathcal{B}_{2\alpha+\beta}$ gives us 
\begin{align}
\text{cov}\mathcal{B}_{0}&=|2p_0(k^{(0)}_0+k^{(0)}_2-k^{(0)}_1-k^{(0)}_3)(l^{(0)}_0-l^{(0)}_1) \nonumber  \\
& +2p_1(k^{(1)}_0+k^{(1)}_2-k^{(1)}_1-k^{(1)}_3)(l^{(1)}_0-l^{(1)}_1) \nonumber  \\
&-\{p_0(k^{(0)}_0+k^{(0)}_2-k^{(0)}_1-k^{(0)}_3) +p_1(k^{(1)}_0+k^{(1)}_2-k^{(1)}_1-k^{(1)}_3)\}  \nonumber  \\
&\times
 \{2p_0 (l^{(0)}_0-l^{(0)}_1)+ 2p_1 (l^{(1)}_0-l^{(1)}_1)\} \nonumber \\
&+ 2p_0(k^{(0)}_0+k^{(0)}_3-k^{(0)}_1-k^{(0)}_2)(l^{(0)}_2-l^{(0)}_3) \nonumber  \\
&+2p_1(k^{(1)}_0+k^{(1)}_3-k^{(1)}_1-k^{(1)}_2)(l^{(1)}_2-l^{(1)}_3) \nonumber  \\
&-\{p_0(k^{(0)}_0-k^{(0)}_2-k^{(0)}_1+k^{(0)}_3) +p_1(k^{(1)}_0-k^{(1)}_2-k^{(1)}_1+k^{(1)}_3)\}  \nonumber  \\
&\times \{2p_0 (l^{(0)}_2-l^{(0)}_3)+ 2p_1 (l^{(1)}_2-l^{(1)}_3)\} |,   
\end{align}

\begin{align}
 \text{cov}\mathcal{B}_{1}&=|2p_0(k^{(0)}_0+k^{(0)}_2-k^{(0)}_1-k^{(0)}_3)(l^{(0)}_2-l^{(0)}_3)  \nonumber  \\
& +2p_1(k^{(1)}_0+k^{(1)}_2-k^{(1)}_1-k^{(1)}_3)(l^{(1)}_2-l^{(1)}_3) \nonumber  \\
&-\{p_0(k^{(0)}_0+k^{(0)}_2-k^{(0)}_1-k^{(0)}_3) +p_1(k^{(1)}_0+k^{(1)}_2-k^{(1)}_1-k^{(1)}_3)\} \nonumber  \\
& \times
 \{2p_0 (l^{(0)}_2-l^{(0)}_3)+ 2p_1 (l^{(1)}_2-l^{(1)}_3)\} \nonumber \\
&+ 2p_0(k^{(0)}_0+k^{(0)}_3-k^{(0)}_1-k^{(0)}_2)(l^{(0)}_0-l^{(0)}_1)  \nonumber  \\
& +2p_1(k^{(1)}_0+k^{(1)}_3-k^{(1)}_1-k^{(1)}_2)(l^{(1)}_0-l^{(1)}_1) \nonumber  \\
&-\{p_0(k^{(0)}_0-k^{(0)}_2-k^{(0)}_1+k^{(0)}_3) +p_1(k^{(1)}_0-k^{(1)}_2-k^{(1)}_1+k^{(1)}_3)\} 
\nonumber  \\
& \times \{2p_0 (l^{(0)}_0-l^{(0)}_1)+ 2p_1 (l^{(1)}_0-l^{(1)}_1)\} |,  
\end{align}
\begin{align}
 \text{cov}\mathcal{B}_{2}&=|2p_0(k^{(0)}_0+k^{(0)}_2-k^{(0)}_1-k^{(0)}_3)(l^{(0)}_0-l^{(0)}_1)  \nonumber  \\
& +2p_1(k^{(1)}_0+k^{(1)}_2-k^{(1)}_1-k^{(1)}_3)(l^{(1)}_0-l^{(1)}_1) \nonumber  \\
&-\{p_0(k^{(0)}_0+k^{(0)}_2-k^{(0)}_1-k^{(0)}_3) +p_1(k^{(1)}_0+k^{(1)}_2-k^{(1)}_1-k^{(1)}_3)\}  \nonumber  \\
& \times
 \{2p_0 (l^{(0)}_0-l^{(0)}_1)+ 2p_1 (l^{(1)}_0-l^{(1)}_1)\} \nonumber \\
&- 2p_0(k^{(0)}_0+k^{(0)}_3-k^{(0)}_1-k^{(0)}_2)(l^{(0)}_2-l^{(0)}_3)  \nonumber  \\
&  -2p_1(k^{(1)}_0+k^{(1)}_3-k^{(1)}_1-k^{(1)}_2)(l^{(1)}_2-l^{(1)}_3) \nonumber  \\
&+\{p_0(k^{(0)}_0-k^{(0)}_2-k^{(0)}_1+k^{(0)}_3) +p_1(k^{(1)}_0-k^{(1)}_2-k^{(1)}_1+k^{(1)}_3)\}  \nonumber  \\
& 
\times \{2p_0 (l^{(0)}_2-l^{(0)}_3)+ 2p_1 (l^{(1)}_2-l^{(1)}_3)\} |,   
\end{align}
\begin{align} 
 \text{cov}\mathcal{B}_{3}&=|2p_0(k^{(0)}_0+k^{(0)}_2-k^{(0)}_1-k^{(0)}_3)(l^{(0)}_2-l^{(0)}_3)  \nonumber  \\
&  +2p_1(k^{(1)}_0+k^{(1)}_2-k^{(1)}_1-k^{(1)}_3)(l^{(1)}_2-l^{(1)}_3) \nonumber  \\
&-\{p_0(k^{(0)}_0+k^{(0)}_2-k^{(0)}_1-k^{(0)}_3) +p_1(k^{(1)}_0+k^{(1)}_2-k^{(1)}_1-k^{(1)}_3)\}  \nonumber  \\
& \times
 \{2p_0 (l^{(0)}_2-l^{(0)}_3)+ 2p_1 (l^{(1)}_2-l^{(1)}_3)\} \nonumber \\
&- 2p_0(k^{(0)}_0+k^{(0)}_3-k^{(0)}_1-k^{(0)}_2)(l^{(0)}_0-l^{(0)}_1) \nonumber  \\
&  -2p_1(k^{(1)}_0+k^{(1)}_3-k^{(1)}_1-k^{(1)}_2)(l^{(1)}_0-l^{(1)}_1) \nonumber  \\
&+\{p_0(k^{(0)}_0-k^{(0)}_2-k^{(0)}_1+k^{(0)}_3) +p_1(k^{(1)}_0-k^{(1)}_2-k^{(1)}_1+k^{(1)}_3)\}  \nonumber  \\
& 
\times \{2p_0 (l^{(0)}_2-l^{(0)}_3)+ 2p_1 (l^{(1)}_2-l^{(1)}_3)\} |. 
\end{align}
Let us assume that the sign of the quantity within the mod of each $\text{cov}\mathcal{B}_{2\alpha+\beta}$ as expressed above is positive. In this case, it can be verified that any of $\Gamma_i$ in $\Gamma$ does not vanish always. It then follows that there are states as in Eq.~(\ref{BLd<=2}) for which $\Gamma>0$. To give an explicit illustration of this, let us consider a state as given by Eq.~(\ref{BLd<=2}) with $p_0=0.4$, $k^{(\lambda)}_i$'s given by 
$k^{(0)}_0=0.4$, $k^{(0)}_1=0.1$, $k^{(0)}_2=0.2$, 
$k^{(1)}_0=0.15$, $k^{(1)}_1=0.15$, $k^{(1)}_2=0.2$
and 
$l^{(\lambda)}_i$'s given by 
$l^{(0)}_0=0.1$,
$l^{(0)}_1=0.4$,
$l^{(0)}_2=0.3$,
$l^{(1)}_0=0.1$,
$l^{(1)}_1=0.1$,
$l^{(1)}_2=0.2$.
For this classical state with $d_\lambda=2$, $\Gamma=0.0192>0$. From this it follows that a nonzero $\Gamma$ cannot be used to witness Bell-local correlations with $d_{\lambda^*}>2$.

\section{Proof of Lemma~\ref{lem2}}\label{dconPR}
We first make the following observation. A convex mixture of any two PR boxes $P^i_{PR}$ and $P^j_{PR}$, here $i=\alpha\oplus\beta \oplus \gamma$ and $j=\alpha'\oplus\beta' \oplus \gamma'$, with $i \ne j$,  can be reexpressed as a convex mixture of a single PR box and a Bell-local distribution with $\Gamma=0$, i.e.,
\be\label{ueqtwoPR}
P=p P^{i}_{PR} + q P^{j}_{PR}= |p-q| P_{NL} + (1-|p-q|) P_{\Gamma=0},
\ee
where  $P_{NL}$ is one of the two PR boxes and $P_{\Gamma=0}=\frac{1}{2}(P^{i}_{PR}+P^{j}_{PR})$, which is a Bell-local distribution since the uniform mixture of any two PR boxes is Bell-local \cite{Bie16}. It is  straightforward to check that the uniform mixture of any two PR boxes also has $\Gamma=0$ and the single PR box fraction $|p-q|$ in Eq.~(\ref{ueqtwoPR}) is related to $\Gamma$ as $|p-q|=\frac{\Gamma(P)}{4}$. As an illustration of this, consider a correlation $P$ which is the uniform mixture of two specific PR boxes given by $P=p P^{000}_{PR}+q P^{111}_{PR}$. For $p>q$, it can be reexpressed as $P=(p-q)P^{000}_{PR}+(1-(p-q))\frac{P^{000}_{PR}+P^{111}_{PR}}{2}$. For the uniform mixture of two PR boxes in this decomposition,  $\Gamma_i$'s are all equal to zero because it has $\text{cov}\mathcal{B}_0=\text{cov}\mathcal{B}_3=1/2$ and $\text{cov}\mathcal{B}_1=\text{cov}\mathcal{B}_2=0$, which implies that $\Gamma=0$, as mentioned in Eq.~(\ref{ueqtwoPR}), and the correlation has $\Gamma(P)=4(p-q)$ because it has $\text{cov}\mathcal{B}_0=p$, $\text{cov}\mathcal{B}_3=q$ and $\text{cov}\mathcal{B}_1=\text{cov}\mathcal{B}_2=0$.

Generalizing Eq.~(\ref{ueqtwoPR}) to the convex mixture of all eight PR boxes, we proceed to prove the lemma.
Let us reexpress the convex combination of eight PR boxes from highest weight to lowest weight as follows: 
\be
\sum_{\alpha\beta\gamma} p_{\alpha\beta\gamma} P^{\alpha\beta\gamma}_{PR}=\sum^{8}_{i=1} q_iP_{PR_i}, 
\ee
where $q_i$'s are one of $p_{\alpha\beta\gamma}$ and satisfy $q_1\ge q_2\ge \cdots \ge q_8$ and $P_{PR_i}$'s are the eight PR boxes having the weights $q_i$'s respectively. Then the convex mixture of the eight PR boxes can be reexpressed as follows:
\be \label{pnlcov8PR}
\sum^{8}_{i=1} q_iP_{PR_i}=p_{NL} P_{PR_1} +(1-p_{NL})P_{L}, 
\ee
where $p_{NL}=(q_1-q_2)-(q_3-q_4)-(q_5-q_6)+(q_7-q_8)$ and $P_{L}$ is a Bell-local distribution given by
\begin{align}\label{PL1}
P_{L}&=\frac{1}{1-p_{NL}}\Big(2q_2\frac{P_{PR_1}+P_{PR_2}}{2}+2q_4\frac{P_{PR_3}+P_{PR_4}}{2}+2q_6\frac{P_{PR_5}+P_{PR_6}}{2} \nonumber \\
&+2q_8\frac{P_{PR_7}+P_{PR_8}}{2}+2(q_3-q_4)\frac{P_{PR_1}+P_{PR_3}}{2} \nonumber \\ 
&+2(q_7-q_8)\frac{P_{PR_5}+P_{PR_7}}{2} +2((q_5-q_6)-(q_7-q_8))\frac{P_{PR_1}+P_{PR_5}}{2} \Big).
\end{align}

Note that $\Gamma_1(\sum_{\alpha,\beta,\gamma} p_{\alpha\beta\gamma} P^{\alpha\beta\gamma}_{PR})=4|\Big||p_{000}-p_{001}|-|p_{010}-p_{011}|\Big|-\Big||p_{100}-p_{101}|-|p_{110}-p_{111}|\Big||$ and similarly, other two  $\Gamma_i(\sum_{\alpha,\beta,\gamma} p_{\alpha\beta\gamma} P^{\alpha\beta\gamma}_{PR})$'s can be expressed. $\Gamma$ of the convex mixture of the eight PR boxes is then given by the minimum between the  $\Gamma_i(\sum_{\alpha,\beta,\gamma} p_{\alpha\beta\gamma} P^{\alpha\beta\gamma}_{PR})$'s. We next note that this minimum is equal to $4p_{NL}$, where $p_{NL}$ is given in Eq.~(\ref{pnlcov8PR}), because this $p_{NL}$ has been minimized overall  possible decompositions into a single PR box and a Bell-local distribution. Using this observation, we have that $P_{L}$ in Eq.~(\ref{PL1}) has $\Gamma=0$.  This follows due to the following reasoning. If $\Gamma(\sum_{\alpha,\beta,\gamma} p_{\alpha\beta\gamma} P^{\alpha\beta\gamma}_{PR})=0$, then $p_{NL}=0$ in the decomposition in Eq.~(\ref{pnlcov8PR}) because $\Gamma(\sum_{\alpha,\beta,\gamma} p_{\alpha\beta\gamma} P^{\alpha\beta\gamma}_{PR})=4p_{NL}$. This holds only if the Bell-local distribution $P_{L}$ has $\Gamma=0$. This completes the proof of Lemma~\ref{lem2}.

%\bibliographystyle{ws-ijqi}
%\bibliography{coherence}

\end{document}